\documentclass[12pt]{article}
\usepackage{epsfig}

\begin{document}

\centerline{\bf \large Pairing in disordered s-wave superconductors 
and}
\centerline{\bf \large the effect of their coupling}
\vspace{0.3cm}

\centerline{Budhaditya Chatterjee$^{1}$, A. Taraphder$^{1,2}$}
\centerline{$^{1}$Department of Physics \& Meteorology} 
\centerline{$^{2}$Centre for Theoretical Studies, IIT Kharagpur 
721302, India}

\begin{abstract}
Inhomogeneity is introduced through random local interactions 
($U_i$) in an attractive
Hubbard model on a square lattice and studied using mean-field 
Bogoliubov-de Gennes 
formalism. Superconductivity is found to get suppressed by the 
random $U_i$     
contrary to the results of a bimodal distribution of $U_i$. The 
proximity effect
of superconductivity is found to be strong, all sites develop 
non-zero 
pairing amplitude. The gap in the density of states is always 
non-zero and 
does not vanish even for strong disorder. When two such 
superconductors are 
coupled via a channel,
the effect of one on the other is negligible. The length 
and width of the
connector, do not seem to have any noticeable effect on the 
superconductivity 
in either systems. The superconducting blocks behave as independent entity and the introduction of
the channel have no effect on them.   
\end{abstract}


\section{Introduction}
One of the most important aspects in the study of correlated system 
is the study of the effect of spatial inhomogeneity\cite{belitz}. 
There have recently
been several examples of such systems where these inhomogeneities 
can 
occur intrinsically via quenched disorder in the system, or can 
occur spontaneously. 
For example, impurities can be driven in a superconductor by 
irradiation
or chemical substitution. On the other hand, holes in the cuprate 
superconductors
or magnetic or charge ordered  domains in manganites spontaneously 
arrange in
geometric patterns like stripes or a checkerboard at certain 
fillings
\cite{mcelroy,mook,hanaguri,tran,renner}. Using scanning tunneling 
microscopy 
evidence for electronic inhomogeneity has been reported in the 
high-Tc 
superconductor $Bi_2Sr_2CaCu_2O_{8+x}$ by McElroy, et al. 
\cite{mcelroy}. Most of the
high-Tc materials are ceramic in nature and inhomogeneities are
present in even the best prepared samples. This inhomogeneity is 
manifested 
as spatial variations in both the local density of states and 
the superconducting energy gap\cite{millis}.

There has been a number of theoretical attempts to understand the 
effect
of quenched disorder in superconductors \cite{dagotto, ghosal}.  
In the context of repulsive models, both in the weak-coupling 
models and their
strong-coupling counterparts, considerable numerical work has been 
done with
inhomogeneities \cite{machida, kato}. In such cases stripes and 
checkerboard
patterns have been reported \cite{white, voita}, the presence of 
d-wave 
superconductivity, however, is less convincing. Enhancement of T$c$ 
due to 
inhomogeneities in the weak-coupling regime is demonstrated by 
Martin et al. 
\cite{martin}. Aryanpour et al. \cite{dagotto} studied an s-wave 
superconductor 
with quenched disorder starting from a negative-U Hubbard model 
using a mean-field 
theory. The disorder enters in their model through a random choice 
of two values 
of the attractive interaction (bimodal distribution) at different 
sites. Quite 
interestingly, it was shown that below a certain value of the 
average attraction, 
the zero 
temperature superconducting gap is larger than that of the 
homogeneous superconductor 
with same (uniform) attraction. Ghosal et al., \cite{ghosal} use an 
s-wave 
superconductor and look for the effect of disorder using a 
mean-field treatment. 
Both these calculation observe strong effects of disorder on the 
superconducting 
order parameters. In certain specific cases, lower dimensional 
orders, like stripes 
or checkerboard order \cite{tran,emery} has been found 
\cite{dagotto} in the 
numerical calculations. One possible application of these 
inhomogeneities is 
to manipulate the degree of superconductivity and the transition 
temperature 
through varying degree of inhomogeneities. Another possibility is 
the study 
of the coupling of two adjacent systems, where superconductivity in
either could be controlled by the degree of disorder in them. In 
many of the
tunneling devices such conditions are obtained.  

We consider a similar situation with the local attractive 
interaction
taken completely random, which, we believe, is more realistic than 
using a
bimodal distribution. We take a system of two such disordered, 
superconducting 
blocks connected by a channel. This model conforms to various 
actual situations 
in tunnel junctions and the study of them is quite 
interesting. We look into the problem of disordered 
superconductivity and compare  
the results with homogeneous systems. Using mean-field exact 
diagonalization 
techniques, we work out the case of random disorder (as opposed to 
forcing 
specific geometrical patterns like stripes\cite{dagotto} from the 
outset) in a 
two dimensional square 
lattice and also for a system of two such blocks with a channel 
connecting them. 
We compute various microscopic parameters such as the local pairing 
amplitude 
$\Delta_i$, the average electron density $n_i$, the local chemical 
potential $\mu_i$ and the density of states for different fillings 
for 
various average interactions. We also look for any emergent one 
dimensional 
pattern in the real-space. The effect of coupling of two such 
systems is
studied in detail. 
 
\section{Model and calculation}

Our starting point is the attractive Hubbard model given by

\begin{equation}
H\ =\ -t \sum_{<ij> \sigma} \big( c^{\dag}_{i \sigma} c_{j \sigma}\ 
+\ c^{\dag}_{j \sigma} c_{i \sigma} \big)\ -\ \mu \sum_{i} 
c^{\dag}_{i
\sigma} c_{i \sigma}\ -\ \sum_{i} |U_{i}|\hat{n_{i \uparrow}} 
\hat{n_{i \downarrow}}
\end{equation} 

\noindent where $t$ is the hopping potential, $\mu$ is the chemical 
potential and $U_i$ 
is the local attractive interaction between the fermion of the 
opposite spins 
residing on the same lattice site $i$. 
$c^\dag_{i\sigma} $ and $c_{i\sigma}$ are creation and destruction 
operators 
for an electron with spin $\sigma$ on site $i$, 
$\hat{n}_{i\uparrow}$ and 
$\hat{n}_{i\downarrow}$ are number operators at site $i$ with spin 
up and down.  
Superconductivity and other charge instabilities in the homogeneous 
case 
have been studied in great detail \cite{at1,at2} in the past. At 
half filling
the superconducting state is degenerate with a charge density wave 
state and
they are connected by a pseudospin rotation.  

In order to study the effect of disorder, an inhomogeneous 
calculation  
is necessary. We use  the Bogoliubov-de Gennes (BdG) mean-field 
approximation
and replace local electron correlation by the local 
superconducting pairing 
amplitude $\Delta_i$, at site $i$ (only s-wave pairing considered 
here). 
$\Delta_i = \langle c_{i\uparrow} c_{i\downarrow} \rangle$ and 
$n_{i\sigma}=
c^\dag_{i\sigma} c_{i\sigma}$. 
Assuming $\langle n_{i \uparrow}\rangle $=$\langle n_{i 
\downarrow}\rangle 
=\langle n_i\rangle /2$
we get,

\begin {equation}
{\cal{H}}_{eff}=\sum_{<ij>,\sigma}(c^{\dag}_{i\sigma}c_{j\sigma} +
c^{\dag}_{j\sigma}c_{i\sigma}) - \sum_{i\sigma}{\tilde\mu_i}
c^{\dag}_{i\sigma}c_{i\sigma}
\nonumber \\
 \sum_{i} 
\big|U_{i}\big|\big[\Delta_{i}c^{\dag}_{i\uparrow}c^{\dag}_
{i\downarrow}+\Delta^{*}_{i}c_{i\downarrow}c_{i\uparrow}\big]\,
\end{equation}

\noindent where $\tilde\mu_i=\mu+ |U_i|\langle n_i\rangle /2 $ is a 
site-dependent 
Hartree shift with $\langle n_i\rangle =\sum_{\sigma}
{\langle n_{i\sigma}\rangle }$. All energies are scaled to $t = 1.$ 
We use an initial value of the chemical potential $\mu$ and use 
different random  
configurations of $U_i$ to realize the inhomogeneous 
superconductor. 
We then compare its tendency of superconductivity with the 
homogeneous system, 
that is, $U_i=U$ for all $i$. For the inhomogeneous case we have 
taken $U_i$ 
to be uniformly random between two specified values. In all these 
cases we use 
the self-consistent diagonalization of the Hamiltonian - at every 
iteration compute 
the averages at every site, recalculate the Hamiltonian and set out 
for diagonalization again. When the iterations
converge we obtain a self-consistent result as usual and then 
compute the values of the physical the quantities.

\section{Results and discussion}

We take a square lattice of finite size and set up the BdG 
Hamiltonian.
On reaching self-consistency, the average electron density, i.e., 
the 
average filling, is calculated. This is repeated for many initial 
$\mu$ and
the results are inverted to find the necessary physical quantities 
for a 
fixed $\langle n\rangle $ as is customary in grand canonical 
ensemble (the local 
inhomogeneity and self-consistent interactions forbid fixing the 
average
electron density). We note that ideally one would expect only 
topological order in two dimensions. However, this does 
not 
apply to mean-field order.  

The variation of $\Delta_i$  with $U_i$ is shown in Fig.1. As 
expected, the
sites with large negative attraction have larger superconducting 
amplitude.
It is interesting to note that although there are sites with 
strictly
zero or very small $|U_i|$, there are no sites with zero 
$\Delta_i$. This
is not unexpected, as a large local fluctuation of order parameter 
is
quite unfavourable with respect to quantum fluctuations and are 
energetically 
costly in terms of the loss of interaction energy. Such finding 
have also
been reported earlier \cite{dagotto}, but we do not find any value 
of 
$\langle U \rangle$ for which the disordered system has a larger 
average
gap than the uniform case. In the case of a bimodal distribution of 
$U_i$,
though, the proximity effect seems to be much stronger and the 
disordered
system has a larger gap than the uniform one below a certain 
$\langle U \rangle$.  
We also observed that unless 
a specific geometry of the disorder is quenched into the lattice, 
there is no 
possibility of a stripe like state to organize self-consistently as 
observed 
by Dagotto\cite{dagotto}. We believe in a situation where the 
disorder
is annealed and randomly distributed, such highly anisotropic 
states are quite 
unlikely in the present model.  
 
\begin{figure}[ht]
\begin{center}
\mbox{\epsfig{file=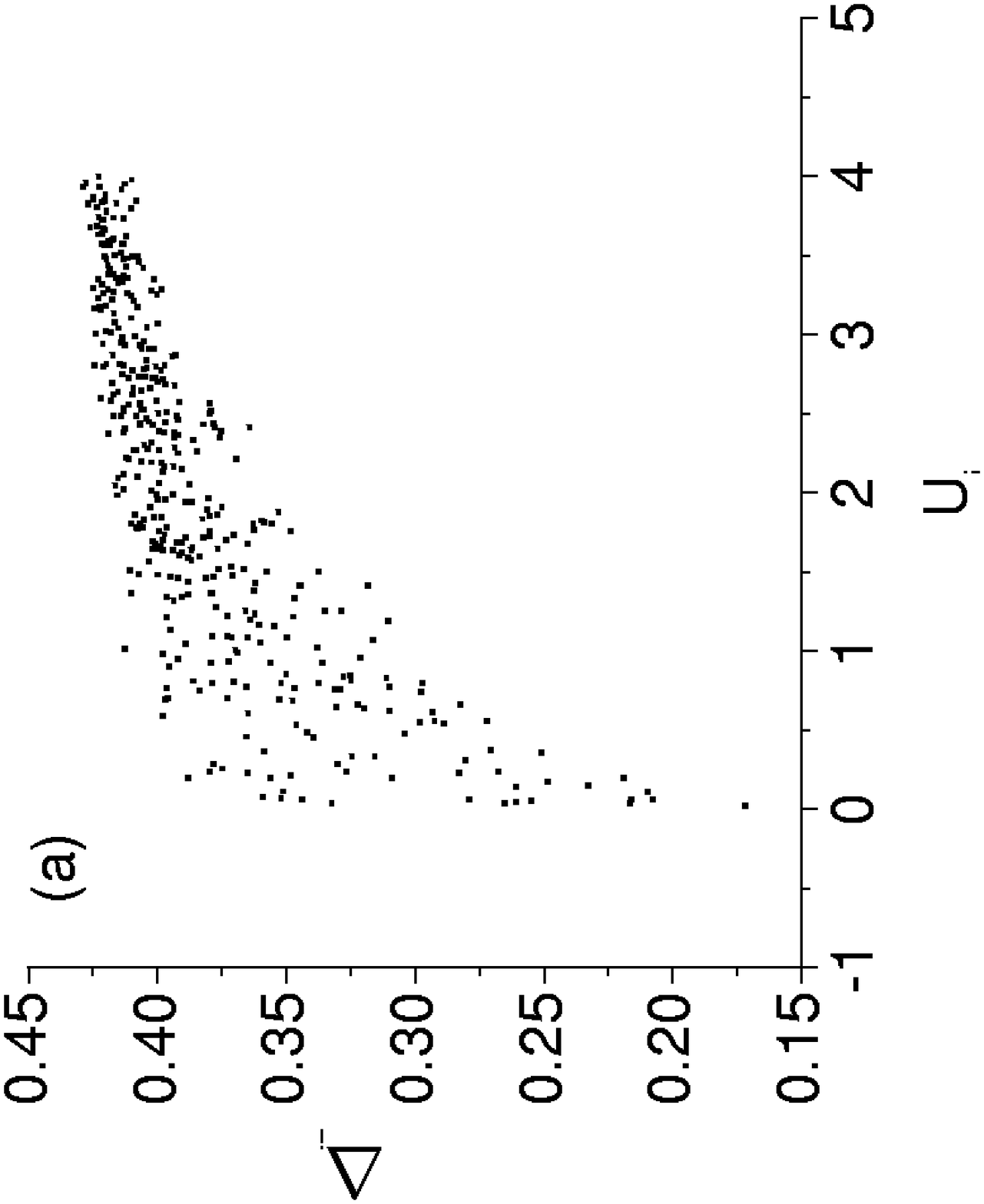,width=1.75in,angle=-90}}
\mbox{\epsfig{file=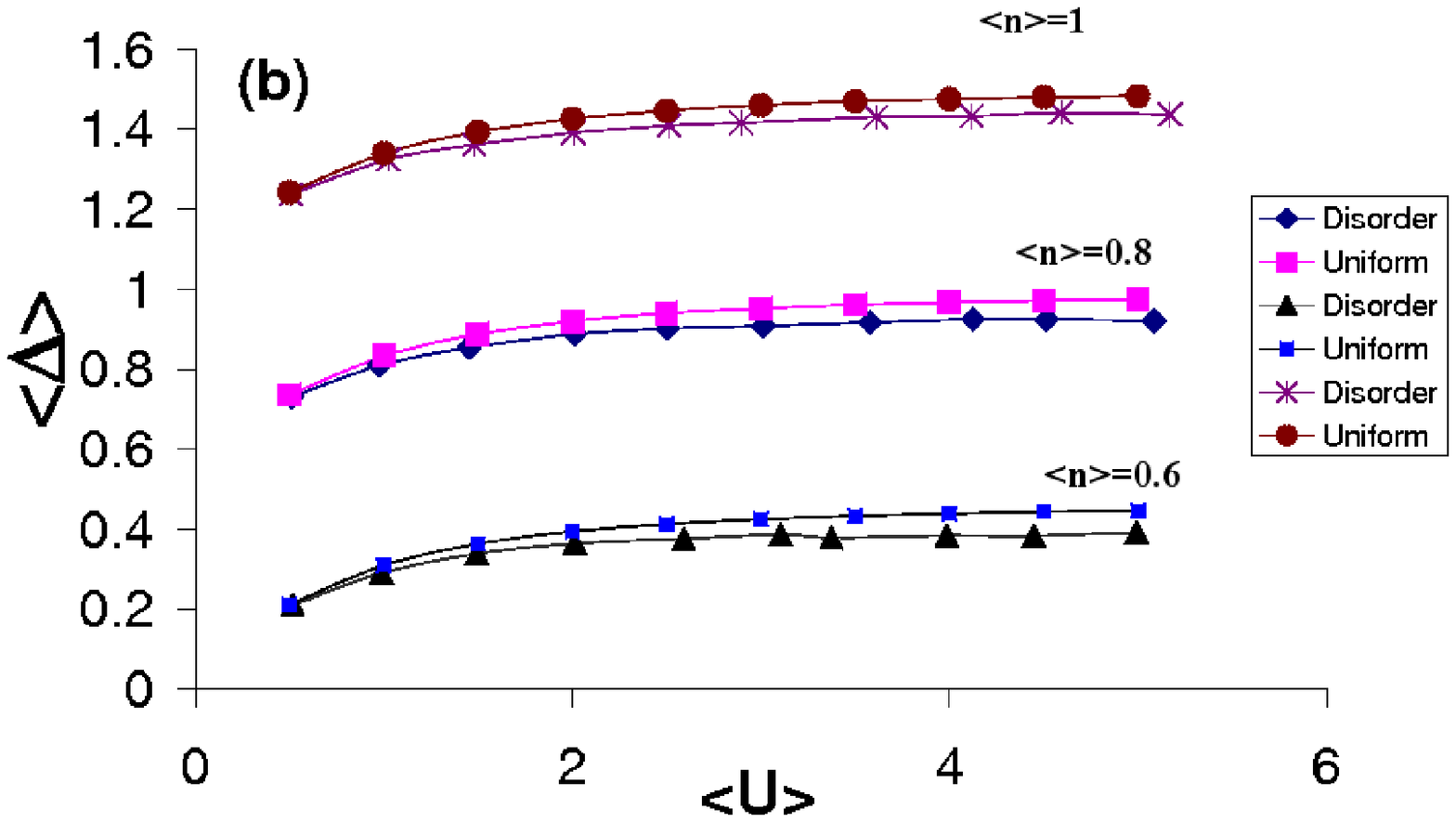,width=1.75in,angle=-90}}
\mbox{\epsfig{file=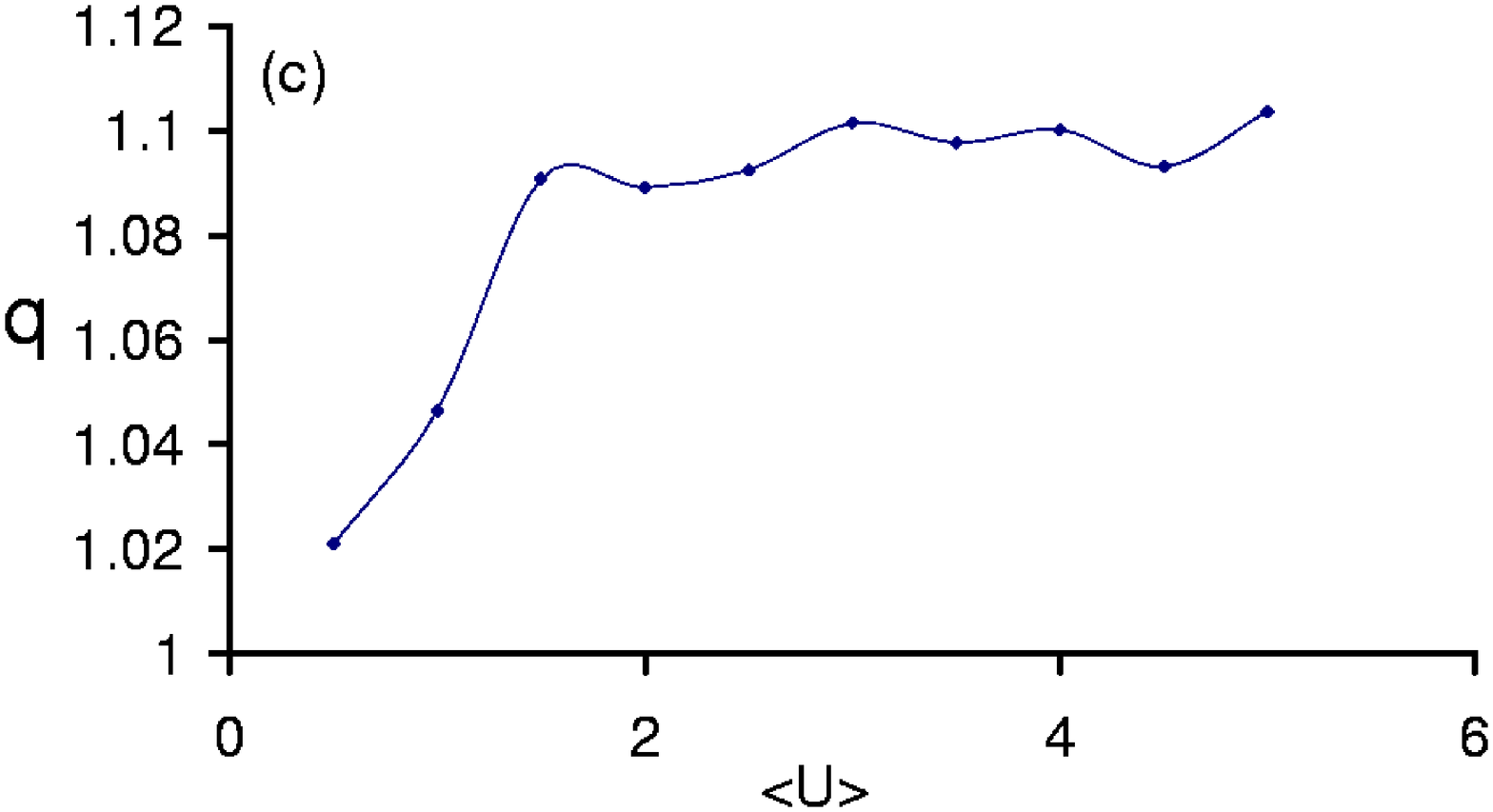,width=1.75in,angle=-90}}
\mbox{\epsfig{file=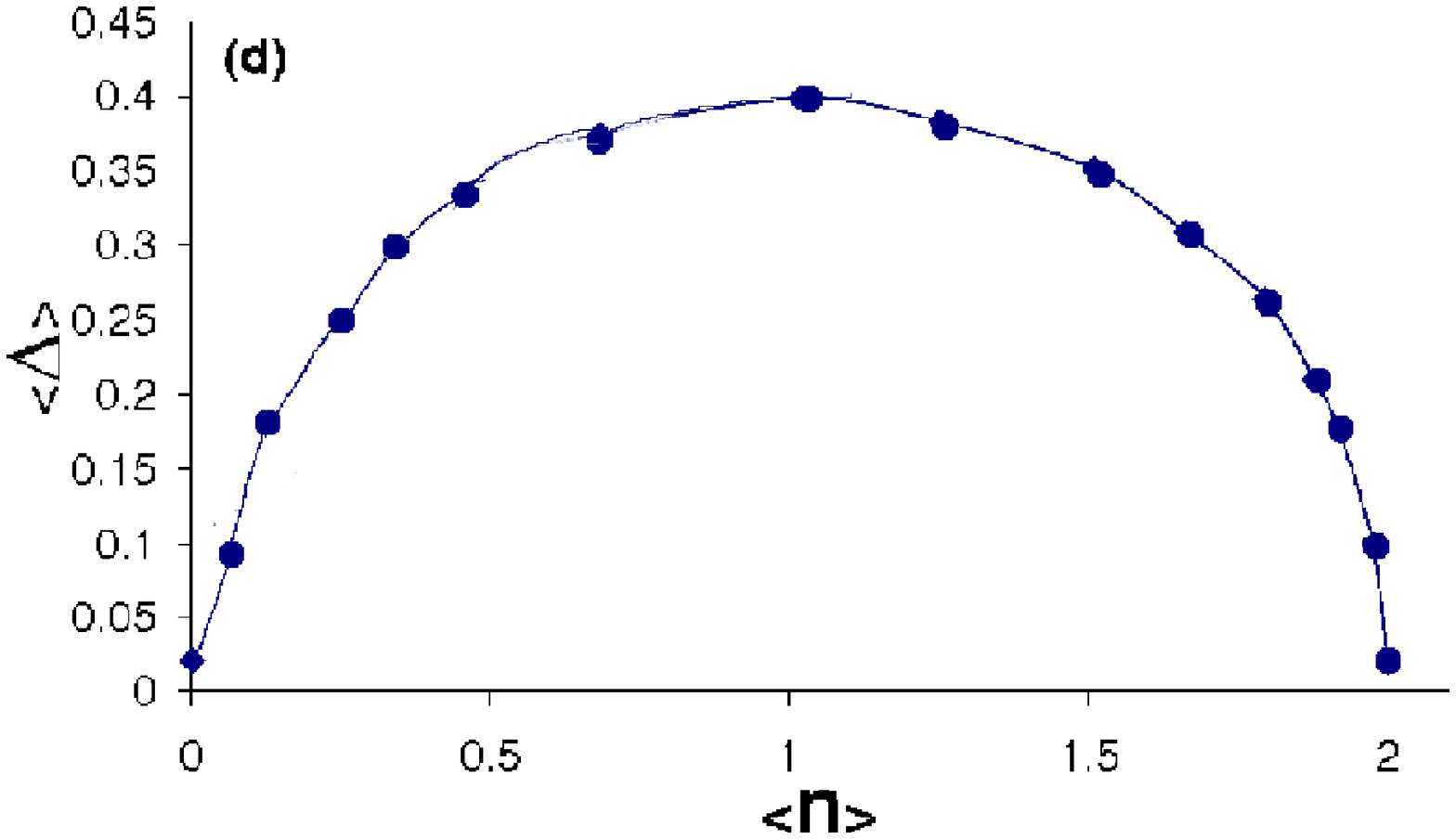,width=1.5in,angle=-90}}
\mbox{\epsfig{file=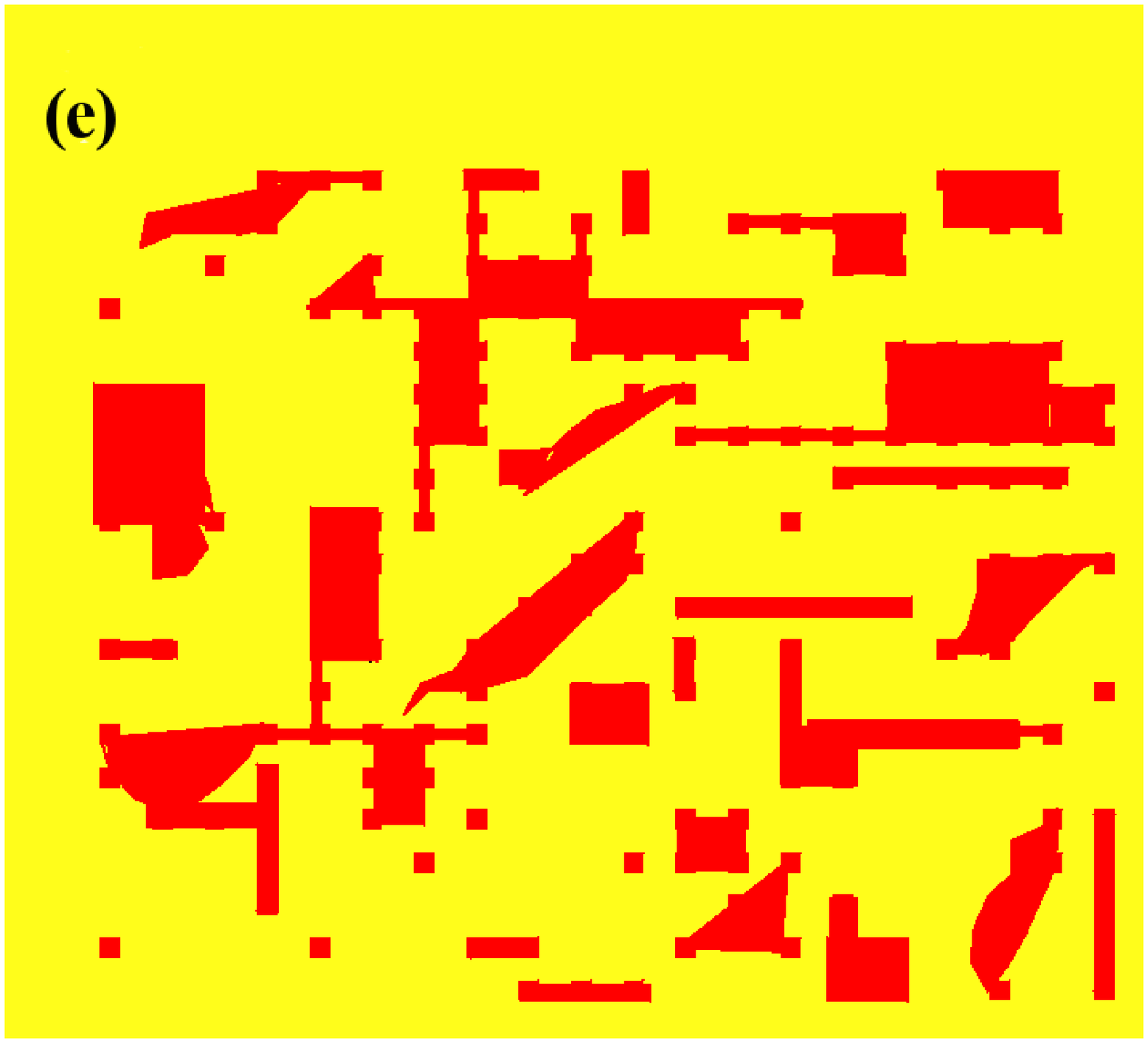,width=1.75in,angle=-90}}
\end{center}
\caption{(Colour online) (a)Variation of the local pairing 
amplitude $\Delta_i$
with the interaction $U_i$, distributed uniformly in the range 0-4
(b)Comparison between the variation of $<\Delta>$ with $<U>$ for 
uniform and disordered cases at $<n>=1$, $<n>=0.8$ and $<n>=0.6$. 
(c)Variation of $<\Delta_{uniform}>/<\Delta_{disorder}> $with 
$<U>$. (d)Variation of average local pairing amplitude$<\Delta>$ 
with $<n>$)(e) Superconducting regions for U=0-2. The red islands 
represent 
superconducting regions with large amplitude (see text).}
\end{figure} 

Fig. 1a shows that $\Delta_i$  increases with $U_i$ and saturates 
at large  $U_i$, there is a broad distribution of $\Delta_i$ 
for
lower values of $U_i$. Fluctuation in the pairing amplitude 
$\Delta_i$ are 
indeed larger in the weak-coupling region. The variation of the 
average $U_i$ and 
the average $<\Delta>= \frac{1}{N}\sum_i \Delta_i$ gives a 
good indication of their relationship and can be compared with the 
uniform case.  
The comparison in Fig.1b shows an enhancement of the average 
superconducting 
pairing amplitude $\langle \Delta \rangle $ with $\langle U \rangle 
$ as 
expected saturating at 
a higher value of the attraction. The value of average pairing 
amplitude is
higher for the uniform case than the disorder case for all values of
$<n>=1.0,0.8$ and $0.6$  contrary to that of bimodal distribution. 
Fig. 1c shows the ratio as
a function of average interaction, and it rises sharply at low $U$ 
and  always
stays above one. This implies that the random disorder inhibits 
long range 
superconducting order. Fig.1d shows the usual bell-shaped curve of $<\Delta>$ versus $<n>$ with 
maximum at $<n>=1$.  

In order to glean the real-space picture of the superconducting
regions, we mark the sites where the value of $\Delta_i$ is greater
than the corresponding value in the homogeneous case (Fig.1e). 
Since all sites
have non-zero $\Delta_i$, this prescription allows us to locate
preferred patterns, if any, in the real space. We do not see any
such pattern while the regions of strong $\Delta_i$ are quite
randomly distributed forming patches. The formation of islands 
in turn reduces the overall observed $T_c$ (which is determined
by the phase coupling across such regions) in a real system and 
the corresponding superfluid stiffness. 

Plotting the frequency distribution of $\Delta_i$ gives a good 
account of the 
effects of disorder. The peak in Fig. 2a is certainly above the 
mean value.  
With higher disorder the peak shifts towards right.  
Further on we calculate the pair correlation function (Fig. 2b), 
defined as  
$c (r_i-r_j)=<\Delta_i\, \Delta_j>$. A disorder averaging is done 
as usual 
to restore the translational invariance.   
The correlation function drops sharply at short distances
but saturates at large distances indicating a true (mean-field) 
long range order.  
The correlation function for four ranges of $U$ are shown in Fig. 
2b, they all
merge to a single curve (except for small fluctuations due to 
finite size) when
normalized. 

\begin{figure}[ht]
\begin{center}
\mbox{\epsfig{file=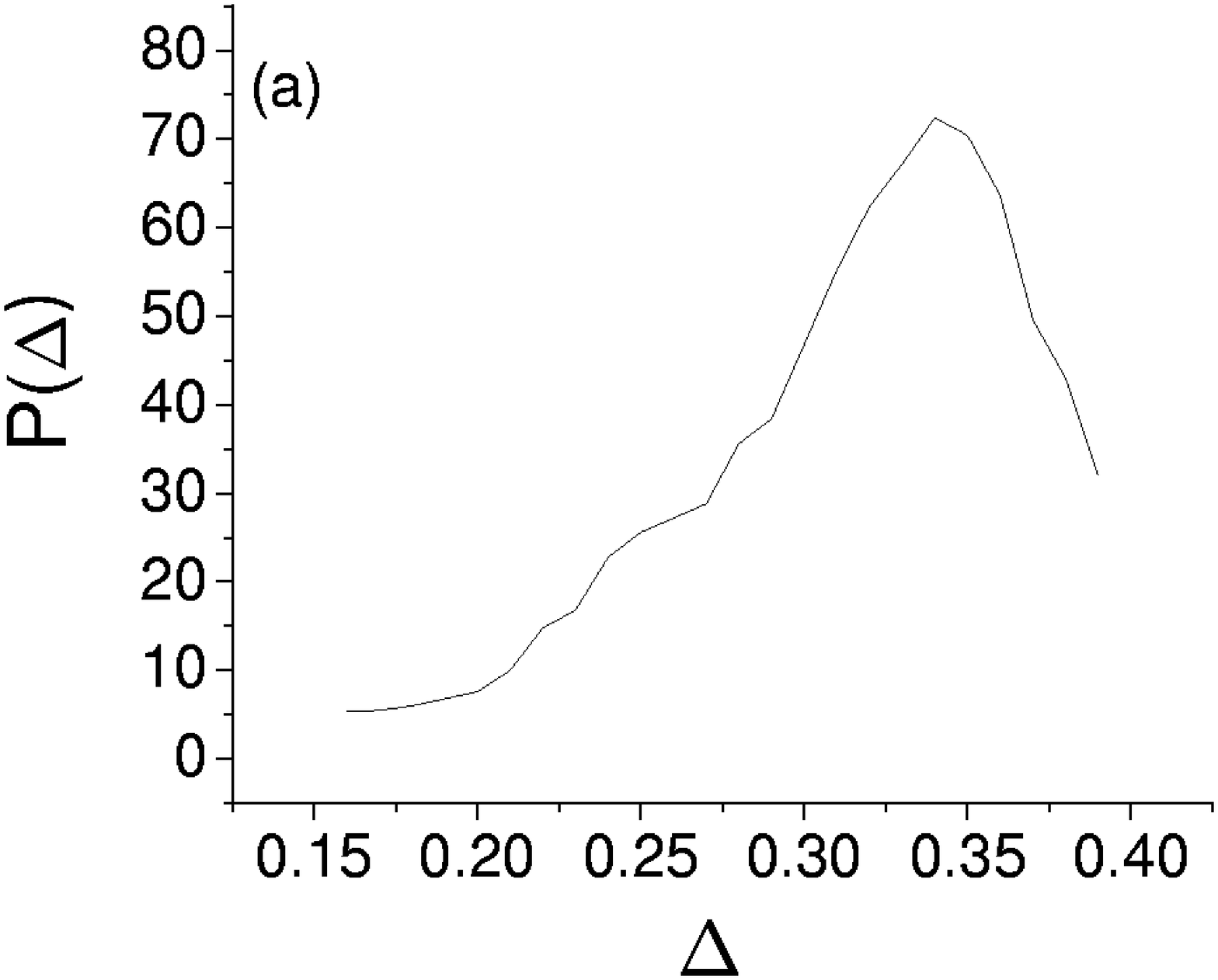,width=1.75in,angle=-90}}
\mbox{\epsfig{file=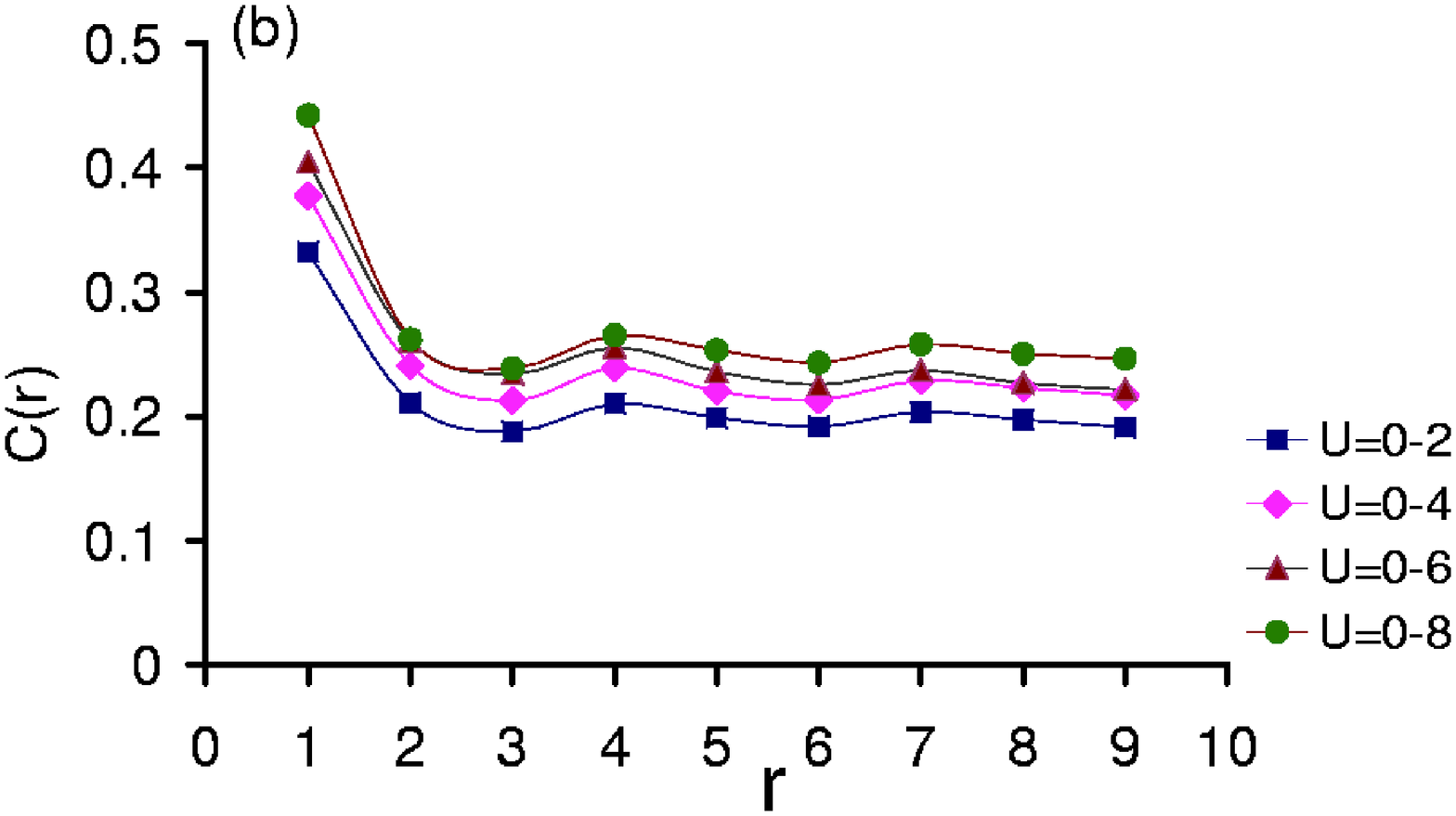,width=1.75in,angle=-90}}
\end{center}
\caption{(a)Frequency Distribution P($\Delta$) of $\Delta$ for 
U=0-2
(b)Plot of correlation for four U ranges: 0-2 0-4 0-6 0-8}
\end{figure}

\begin{figure}[ht]
\begin{center}
\mbox{\epsfig{file=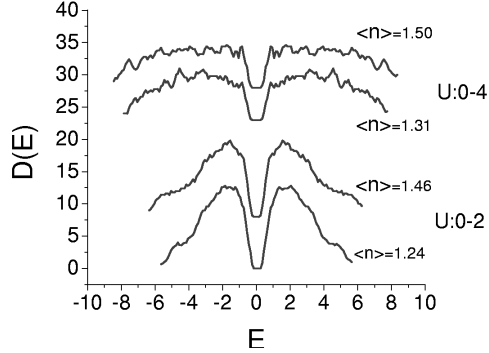,width=2in,angle=-90}}
\end{center}
\caption{Density of States plot for U=0-2 and 0-4, the two sets of 
curves are 
offset vertically to make them separate.}
\end{figure}

The presence of a superconducting gap is another indication of the 
true long
range order \cite{ghosal}. It is, therefore, very important to look 
at the density 
of states (DOS) of the system. As mentioned above there are no 
sites with zero 
pairing amplitude. This is reflected in the DOS where the gap at 
zero energy
is clearly seen in Fig. 3, while the very sharp peaks (divergences 
in a  
homogeneous superconductor) are now broadened into 
two symmetrical regions with a broad distribution of states at 
higher and lower 
energies. The presence of the energy gap even in the highly 
disordered systems 
clearly indicates that s-wave superconductivity is possible even in 
the presence of large disorder.  

\begin{figure}[ht]
\begin{center}
       \epsfig{file=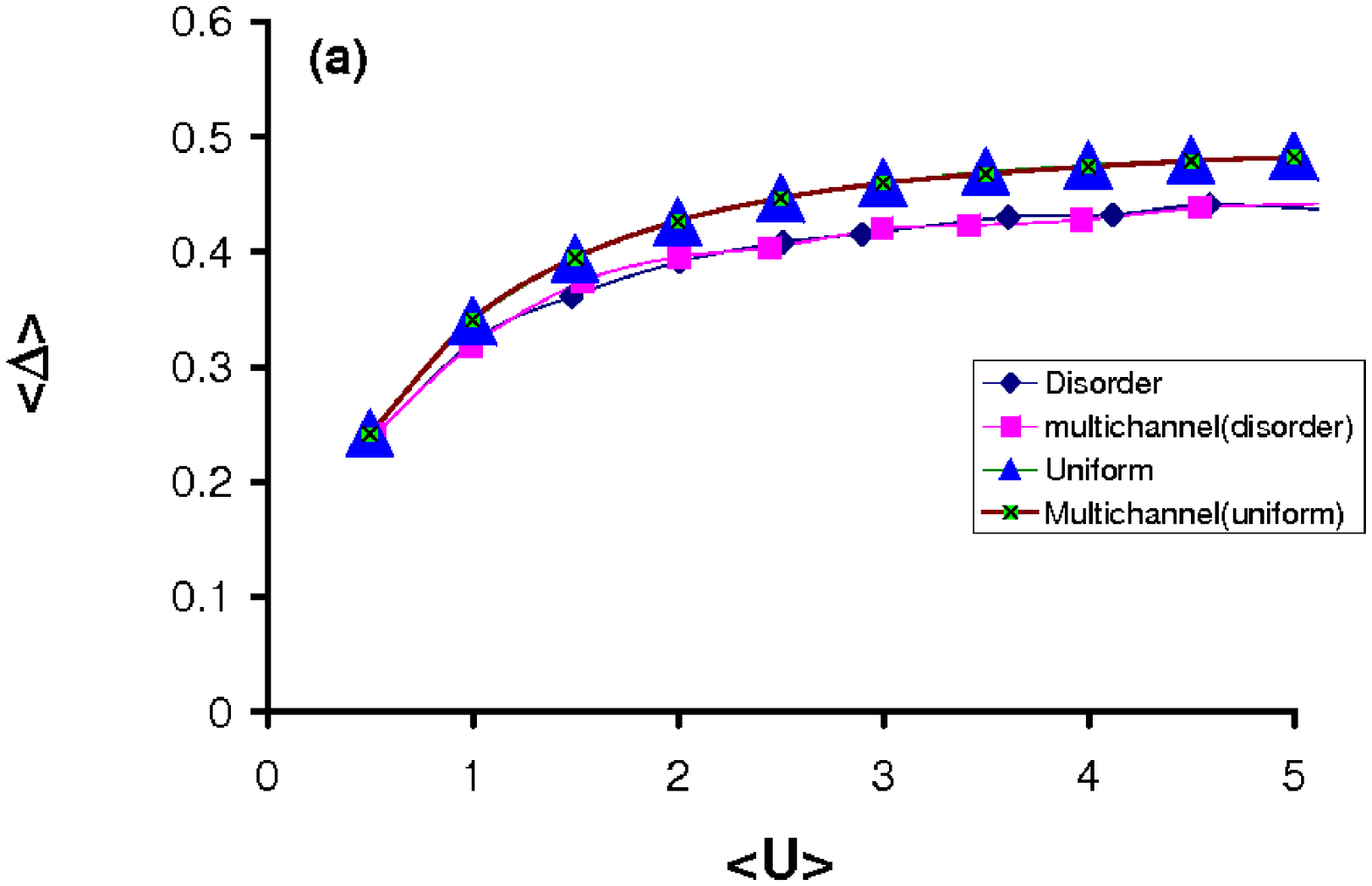,width=1.75in,angle=-90}
       \epsfig{file=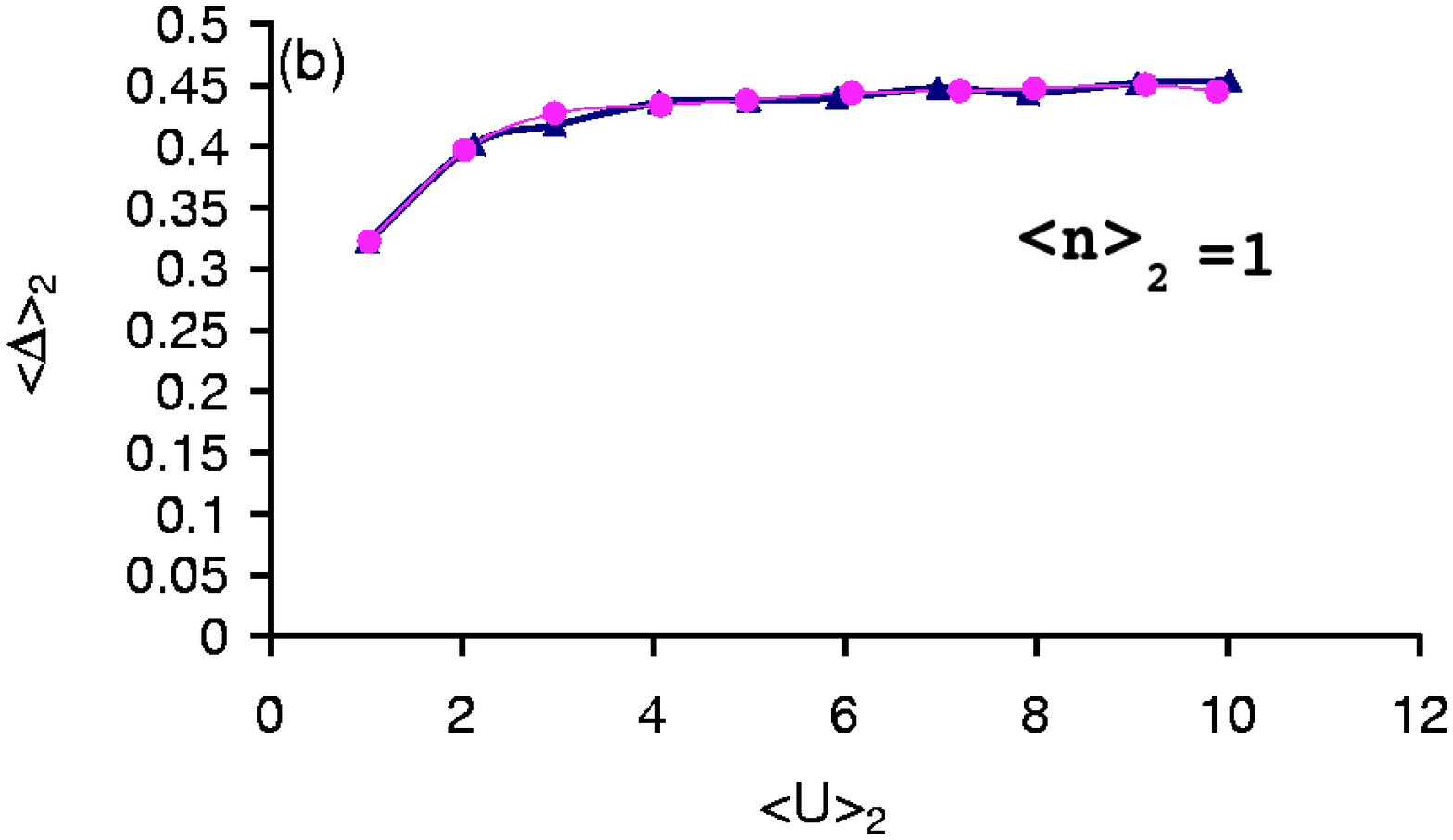,width=1.75in,angle=-90}
       \epsfig{file=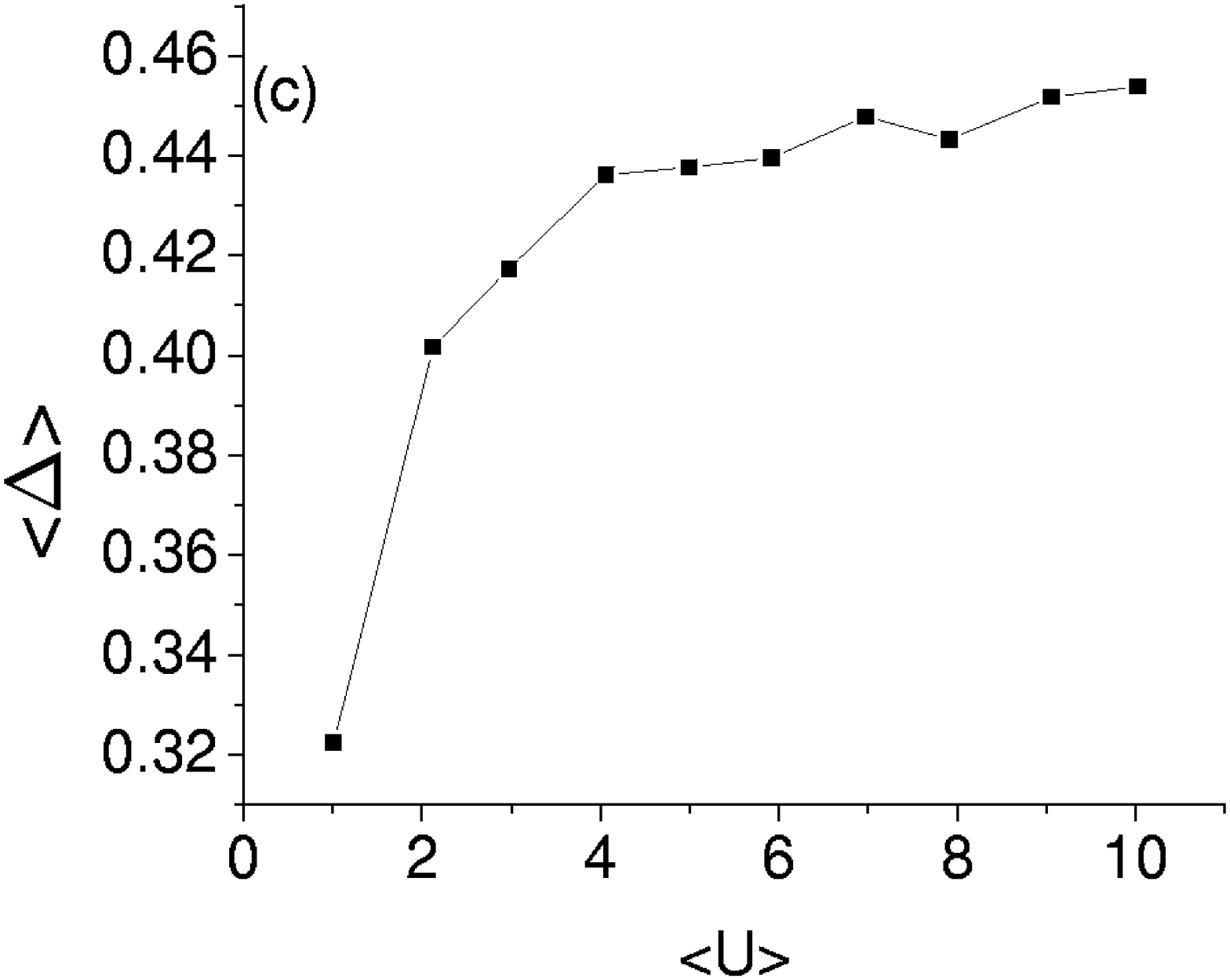,width=1.75in,angle=-90}
       \epsfig{file=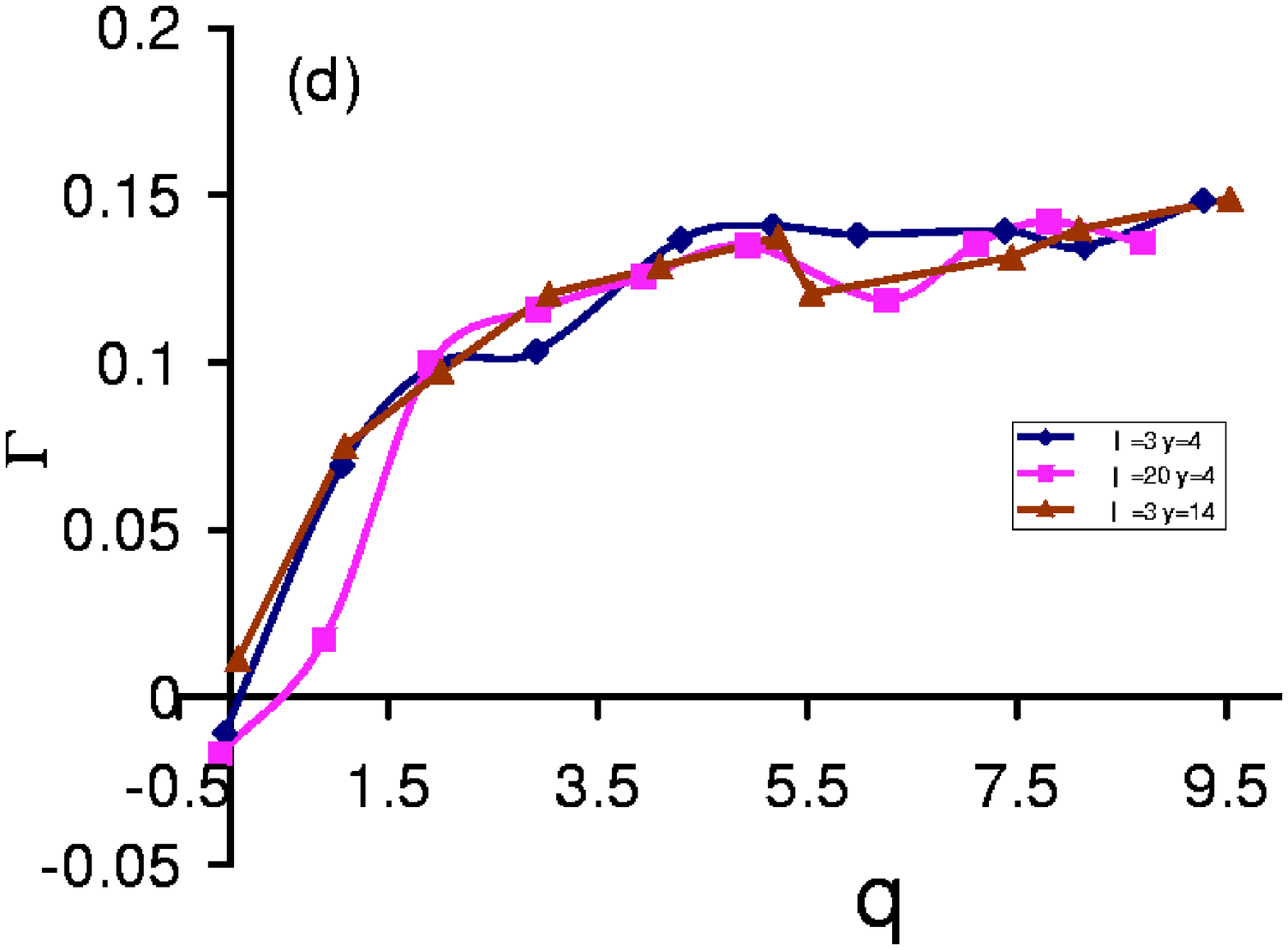,width=1.75in,angle=-90}
       \epsfig{file=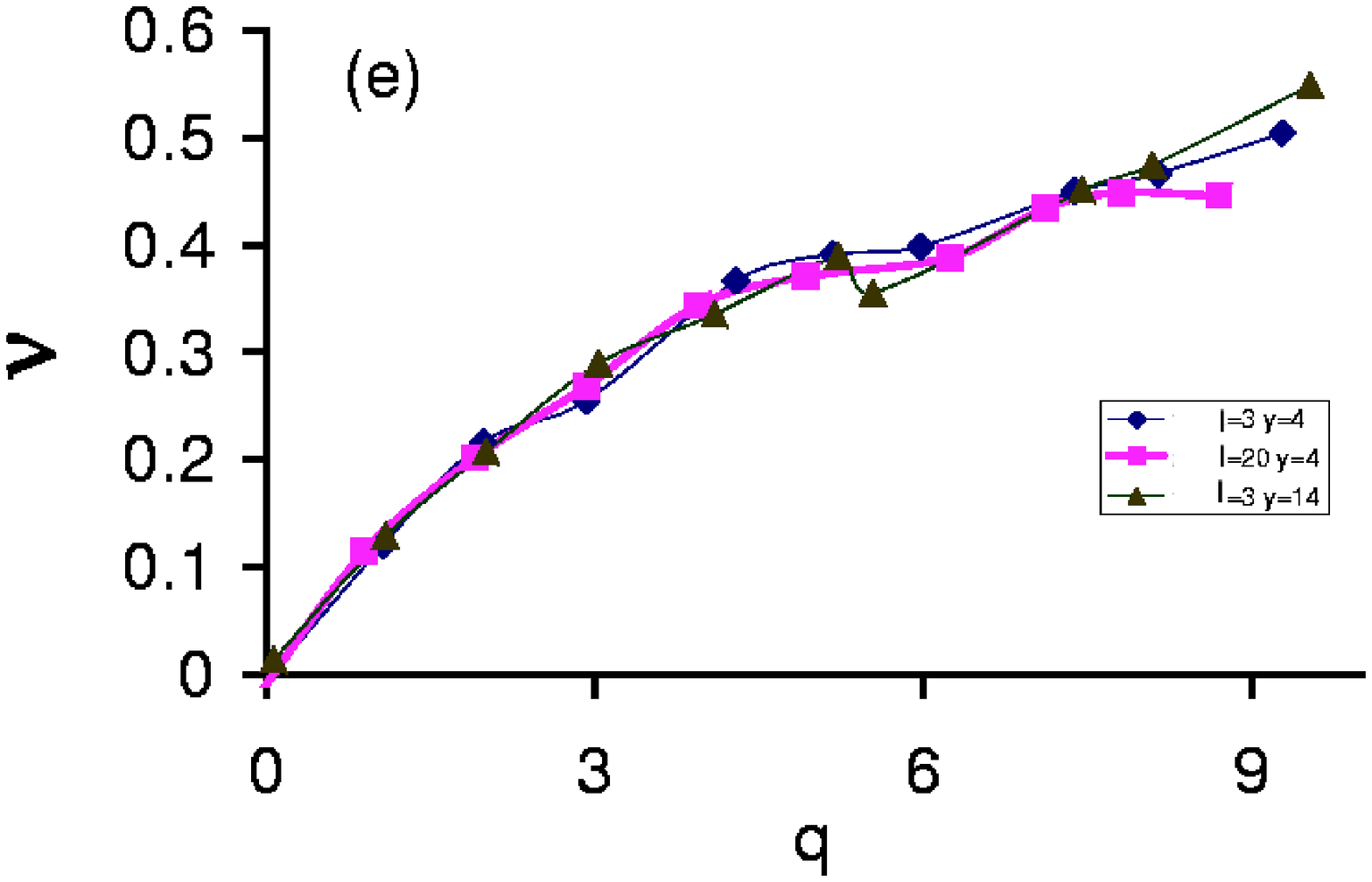,width=1.9in,angle=-90}
       \epsfig{file=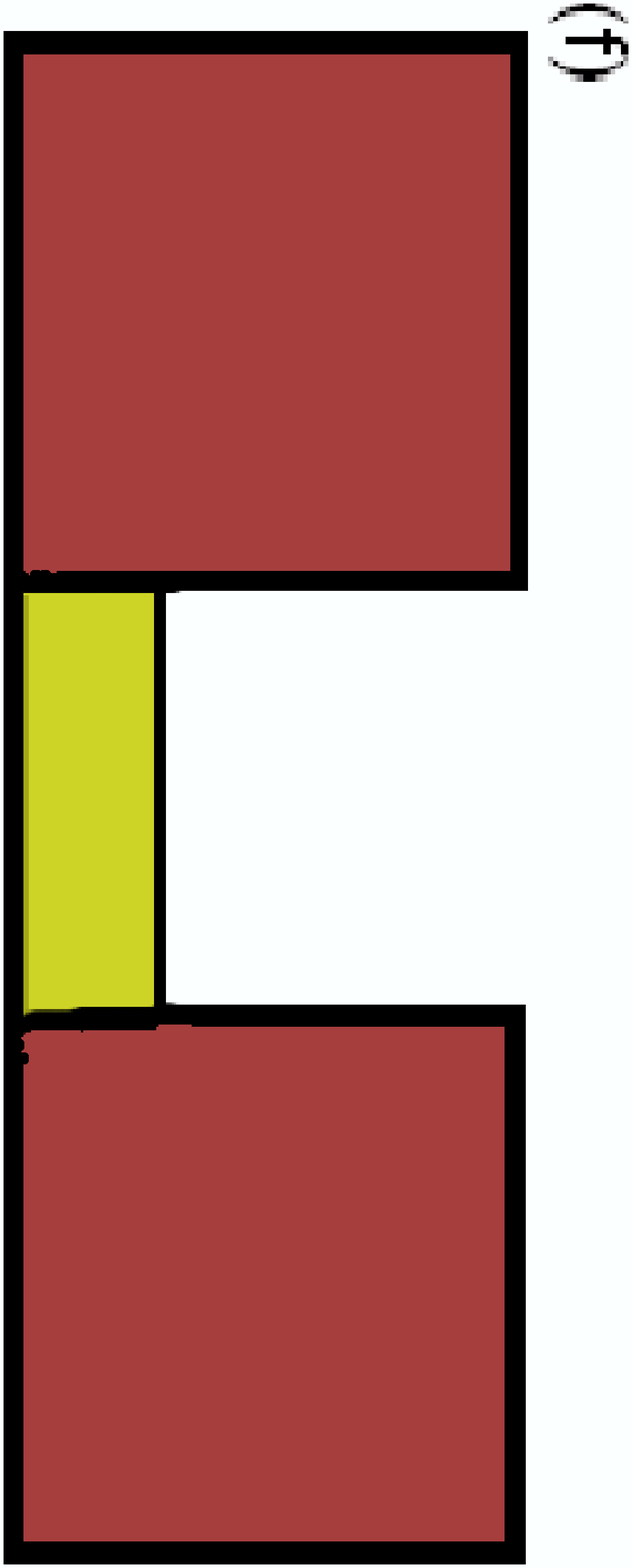,width=1.9in,angle=90}
\end{center}
\caption{ (a)Variation of $<\Delta>$ with $<U>$ for individual blocks of the coupled system for 
both disorder and uniform case and its comparison with single block system.  
(b) $\langle \Delta\rangle_2$ as a function of $\langle U\rangle_2 
$ 
for $\langle n\rangle_2=1$ and comparison with superconductivity in 
a single system.
(c) $\langle \Delta\rangle $ as a function of $\langle U\rangle $ 
with $\langle n\rangle =1$ for the whole system. (d) Variation 
of $ \Gamma= \langle \Delta\rangle_2 - \langle \Delta\rangle_1 $ 
with $ q= \langle U\rangle_2 -\langle U\rangle_1 ,\,   \langle 
n\rangle =1$. Here
the legend $l$ refers to length of channel and $y$ refers to its 
width.   
(e) Variation of $\nu= \langle n\rangle_2 - \langle n\rangle_1$ 
with  $ q $ while $ \langle n\rangle =1.$
(f) Two superconducting blocks connected by a channel} 
\end{figure}

To understand the effect of coupling of two disordered 
superconductors, we take up 
two such superconductors and join them via a narrow
channel (Fig. 4f) and observe the effects due to the proximity of 
each other.   
We find that contrary to the common expectation the width and length of the channel do 
not have any significant effect on the superconductivity in
either systems (4d,e).More interestingly, results from the 
comparative variation of $\langle \Delta\rangle, \langle U\rangle, 
\langle n\rangle $ of the two blocks (Fig. 4a-e) do not give any clear influence of the channel.
 For example, we 
observed the variation of $\langle\Delta\rangle$ of one 
superconductor 
with  the $ \langle U\rangle $ in both uniform and disordered case and compared with  similar results of the single block system
 we had obtained earlier. Results are 
symmetric 
with respect
to interchange of systems 1 and 2. It is clearly seen from Fig. 
4a that
the average superconducting pairing amplitude in blocks of the coupled system follow identically 
the pattern of a single block system in both disorder and uniform regime. 
 The 
channel was kept 
homogeneous. The results do not change if the channel is maintained 
homogeneous but non superconducting ($U_i=0$ for all $i$ in the channel).This unexpected behaviour,  
we believe, is due to the formation of islands  of low  pairing amplitude discussed  earlier (Fig 1e).
These regions localize the 
electron in the block since it would be energetically unfavourable for the electrons to percolate through the channel.
 This inhibits correlations between the blocks,effectively making them behave as  independent
systems.

The average superconducting pairing amplitude in the 
second 
system increases at the expense of 
$\langle \Delta_1\rangle $ (Fig. 4b since 
the enhanced attraction in system 2 provides extra stabilization 
energy
in that region. Note that the variation of $<\Delta_2>$ with 
$<U_2>$ is
also  unaffected (Fig. 4b) by the presence of system 1 and nearly 
similar to that  for a single system. The variations of overall 
average 
pairing amplitude $\langle \Delta_1 + \Delta_2 \rangle$ with total 
average
interaction at a fixed overall density is shown in Fig. 4c. The 
combined
system behaves like an isolated single system  and expectedly 
shows the
typical rise and saturation behaviour seen in a single system.    

Fig. 4d demonstrates clearly how the average order parameter in 
the two systems
are correlated with the corresponding interactions. The larger the 
attraction,
more is the average $ \langle \Delta \rangle $. As  $ \langle U 
\rangle $ increases
in any system, the density of electrons tends to increase for the 
extra stabilization available in that region. Fig. 4e shows this 
tendency,
with  $ \langle n_i \rangle $ in each superconductor increasing 
rapidly with
the average attractive interaction there.  
 
\section{Conclusion}

In conclusion, we have studied inhomogeneous s-wave superconductors 
using 
the Bogoliubov de-Gennes mean-field theory. Superconductivity is 
suppressed 
over the homogeneous weak-coupling value due to disorder, though 
the proximity
effect is strong with non-zero pairing amplitude at all sites 
(even with
$U_i=0$). The frequency distribution of 
$ p(\Delta)$ 
shows a peak towards higher values of $\Delta$. The gap in the 
density of states 
persists even for the high disordered case lending support to the
existence of strongly disordered inhomogeneous superconductivity. 
When two such blocks of inhomogeneous superconductors are connected 
via a
channel, there is no appreciable effect of one on the other.
 The individual blocks are not affected significantly by the other block connected by the channel .
We argue that this is due to the fact that the fluctuating local order parameters remain pinned 
to the individual values thereby preventing significant correlation between the blocks.

\section{Acknowledgment}
BC acknowledges a junior research fellowship from the Council
of Scientific and Industrial Research, India.

\end{document}